\documentclass[11pt,a4paper]{article}
\usepackage[hyperref]{acl2019}
\usepackage{times}
\usepackage{latexsym}
\usepackage{xcolor}
\usepackage{url}
\usepackage{algorithm2e}
\usepackage{graphicx}
\usepackage{amsmath}

\aclfinalcopy 

\title{Two-stage Federated Phenotyping and Patient Representation Learning}

\author{Dianbo Liu  \\
    CHIP\\
  Boston Children's Hospital \\ Harvard Medical School\\Boston, MA, USA, 02115 \\
  \text{dianbo.liu@childrens.harvard.edu} \\\And
  Dmitriy Dligach \\
 Loyola University Chicago \\ Chicago, IL, USA 60660 \\
  \text{ddligach@luc.edu} \\\And
  Timothy Miller \\
  CHIP\\
   Boston Children's Hospital \\ Harvard Medical School\\Boston, MA, USA, 02115 \\
  \text{timothy.miller@childrens.harvard.edu} \\}

\date{}

\begin{document}
\maketitle
\begin{abstract}
 A large percentage of medical information is in unstructured text format in electronic medical record systems. Manual extraction of information from clinical notes is extremely time consuming. Natural language processing has been widely used in recent years for automatic information extraction from medical texts. However, algorithms trained on data from a single healthcare provider are not generalizable and error-prone due to the heterogeneity and uniqueness of medical documents. We develop a two-stage federated natural language processing method that enables utilization of clinical notes from different hospitals or clinics without moving the data, and demonstrate its performance using obesity and comorbities phenotyping as medical task. This approach not only improves the quality of a specific clinical task but also facilitates knowledge progression in the whole healthcare system, which is an essential part of learning health system. To the best of our knowledge, this is the first application of federated machine learning in clinical NLP.
\end{abstract}

\section{Introduction}

Clinical notes and other unstructured data in plain text are valuable resources for medical informatics studies and machine learning applications in healthcare. In clinical settings, more than 70\% of information are stored as unstructured text. Converting the unstructured data into useful structured representations will not only help data analysis but also improve efficiency in clinical practice \cite{jagannathan2009assessment,kreimeyer2017natural,ford2016extracting,demner2009can,murff2011automated,friedman2004automated}. Manual extraction of information from the vast volume of notes from electronic health record (EHR) systems is too time consuming. 

To automatically retrieve information from unstructured notes, natural language processing (NLP) has been widely used. NLP is a subfield of computer science, that has been developing for more than 50 years, focusing on intelligent  processing of human languages \cite{manning1999foundations}. A combination of hard-coded rules and machine learning methods have been used in the field, with machine learning currently being the dominant paradigm.


Automatic phenotyping is a task in clinical NLP that aims to identify cohorts of patients that match a predefined set of criteria. Supervised machine learning is curently the main approach to phenotyping, but availability of annotated data hinders the progress for this task. In this work, we consider a scenario where multiple instituitions have access to relatively small amounts of annotated data for a particular phenotype and this amount is not sufficient for training an accurate classifier. On the other hand, combining data from these institutions can lead to a high accuracy classifier, but direct data sharing is not possible due to operational and privacy  concerns. 

Another problem we are considering is learning patient representations that can be used to train accurate phenotyping classifiers. The goal of patient representation learning is mapping the text of notes for a patient to a fixed-length dense vector (embedding). Patient representation learning has been done in a supervised \cite{dligach2018learning} and unsupervised \cite{miotto2016deep} setting. In both cases, patient representation learning requires massive amounts of data. As in the scenario we outlined in the previous paragraph, combining data from several institutions can lead to higher quality patient representations, which in turn will improve the accuracy of phenotyping classifiers. However, direct data sharing, again, is difficult or impossible.


To tackle the challenges we mentioned above, we developed a federated machine learning method to utilize clinical notes from multiple sources, both for learning patient representations and phenotype classifiers. 


Federated machine learning is a concept that machine learning models are trained in a distributed and collaborative manner without centralised data~\cite{liu2018fadl,mcmahan2016communication,bonawitz2019towards,konevcny2016federated,huang2018loadaboost,huang2019patient}. The strategy of federated learning has been recently adopted in the medical field in structured data-based machine learning tasks~\cite{liu2018fadl,huang2018loadaboost,liu2018artificial}. However, to the best of our knowledge, this work is the first time a federated learning strategy has been used in medical NLP. 

We developed our two-stage federated natural language processing method based on previous work on patient representation \cite{dligach2018learning}. The first stage of our proposed federated learning scheme is supervised patient representation learning. Machine learning models are trained using medical notes from a large number of hospitals or clinics without moving or aggregating the notes. The notes used in this stage need not be directly relevant to a specific medical task of interest. At the second stage, representations from the clinical notes directly related to the phenotyping task are extracted using the algorithm obtained from stage 1 and a machine learning model specific to the medical task is trained.


Clinicians spend a significant amount of time reviewing clinical notes. This time can be saved or reduced with reasonably designed NLP technologies. One such task is phenotying from medical notes. In this study, we demonstrated, using phenotyping from clinical note as a clinical task \cite{conway2011analyzing,dligach2018learning}, that the method we developed will make it possible to utilize notes from a wide range of hospitals without moving the data. 

The ability to utilize clinical notes distributed at different healthcare providers not only benefits a specific clinical practice task but also facilitates building a learning healthcare system, in which meaningful use of knowledge in distributed clinical notes will speed up progression of medical knowledge to translational research, tool development, and healthcare quality assessment \cite{friedman2010achieving,blumenthal2010meaningful}. Without the needs of data movement, the speed of information flow can approach real time and make a rapid learning healthcare system possible \cite{slutsky2007moving,friedman2014toward,abernethy2010rapid}. 

\section{Methods}

\subsection{Study Cohorts}

Two datasets were used in this study. The MIMIC-III corpus~\cite{johnson2016mimic} was used for representation learning. This corpus contains information for more than 58,000 admissions for more than 45,000 patients admitted to Beth Israel Deaconess Medical Center in Boston between 2001 and 2012. Relevant to this study, MIMIC-III includes clinical notes, ICD9 diagnostic codes, ICD9 procedure codes, and CPT codes. The notes were processed with cTAKES\footnote{\url{https://ctakes.apache.org}} to extract UMLS\footnote{\url{https://www.nlm.nih.gov/research/umls/}} unique concept identifiers (CUIs). Following the cohort selection protocol from \cite{dligach2018learning}, patients with over 10,000 CUIs were excluded from this study. We obtained a cohort of 44,211 patients in total. 

The Informatics for Integrating Biology to the Bedside (i2b2) Obesity challenge dataset was used to train phenotyping models~\cite{uzuner2009recognizing}. The dataset consists of 1237 discharge summaries from Partners HealthCare in Boston. Patients in this cohort were annotated with respect to obesity and its comorbidities. In this study we consider the more challenging \textit{intuitive} version of the task. The discharge summaries were annotated with obesity and its 15 most common comorbidities, the presence, absence or uncertainty (questionable) of which were used as ground truth label in the phenotyping task in this study. Table \ref{i2b2Cohort} shows the number of examples of each class for each phenotype. Thus, we build phenotyping models for 16 different diseases.

\begin{table*}[t!]
\centering 
\caption{i2b2 cohort of obesity comorbidities}

\begin{tabular}{|l|c|c|c|}

\hline
\textbf{Disease}              & \textbf{\#Absence} & \textbf{\#Presence} & \textbf{\#Questionable} \\ \hline
\textbf{Asthma}               & 86                 & 596                 & 0                       \\ \hline
\textbf{CAD}                  & 391                & 265                 & 5                       \\ \hline
\textbf{CHF}                  & 308                & 318                 & 1                       \\ \hline
\textbf{Depression}           & 142                & 555                 & 0                       \\ \hline
\textbf{Diabetes}             & 473                & 205                 & 5                       \\ \hline
\textbf{GERD}                 & 144                & 447                 & 1                       \\ \hline
\textbf{Gallstones}           & 101                & 609                 & 0                       \\ \hline
\textbf{Gout}                 & 94                 & 616                 & 2                       \\ \hline
\textbf{Hypercholesterolemia} & 315                & 287                 & 1                       \\ \hline
\textbf{Hypertension}         & 511                & 127                 & 0                       \\ \hline
\textbf{Hypertriglyceridemia} & 37                 & 665                 & 0                       \\ \hline
\textbf{OA}                   & 117                & 554                 & 1                       \\ \hline
\textbf{OSA}                  & 99                 & 606                 & 8                       \\ \hline
\textbf{Obesity}              & 285                & 379                 & 1                       \\ \hline
\textbf{PVD}                  & 110                & 556                 & 1                       \\ \hline
\textbf{Venous Insufficiency} & 54                 & 577                 & 0                       \\ \hline
\end{tabular}
\label{i2b2Cohort}
\end{table*}

\subsection{Data Extraction and feature choice}
At the representation learning stage (stage 1), all notes for a  patient were aggregated into a single document. CUIs extracted from the text were used as input features. ICD-9 and CPT codes for the patient were used as labels for supervised representation learning.

At the phenotyping stage (stage 2), CUIs extracted from the discharge summaries were used as input features. Annotations of being present, absent, or questionable for each of the 16 diagnoses for each patient were used as multi-class classification labels.

\subsection{Two-stage federated natural language processing of clinical notes}

We envision that clinical textual data can be useful in at least two ways: (1) for pre-training patient representation models, and (2) for training phenotyping models. 

In this study, a patient representation refers to a fixed-length vector derived from clinical notes that encodes all essential information about the patient. A patient representation model trained on massive amounts of text data can be useful for a wide range of clinical applications. A phenotyping model, on the other hand, captures the way a specific medical condition works, by learning the function that can predict a disease (e.g., asthma) from the text of the notes. 

Until recently, phenotyping models have been trained from scratch, omitting stage (1), but recent work~\cite{dligach2018learning} included a pre-training step, which derived dense patient representations from data linking large amounts of patient notes to ICD codes. Their work showed that including the pre-training step led to learning patient representations that were more accurate for a number of phenotyping tasks.

Our goal here is to develop methods for federated learning for both (1) pre-training patient representations, and (2) phenotyping tasks. These methods will allow researchers and clinicans to utilize data from multiple health care providers, without the need to share the data directly, obviating issues related to data transfer and privacy.

\begin{figure*}[t!]
  \centering 
  \includegraphics[width=5in]{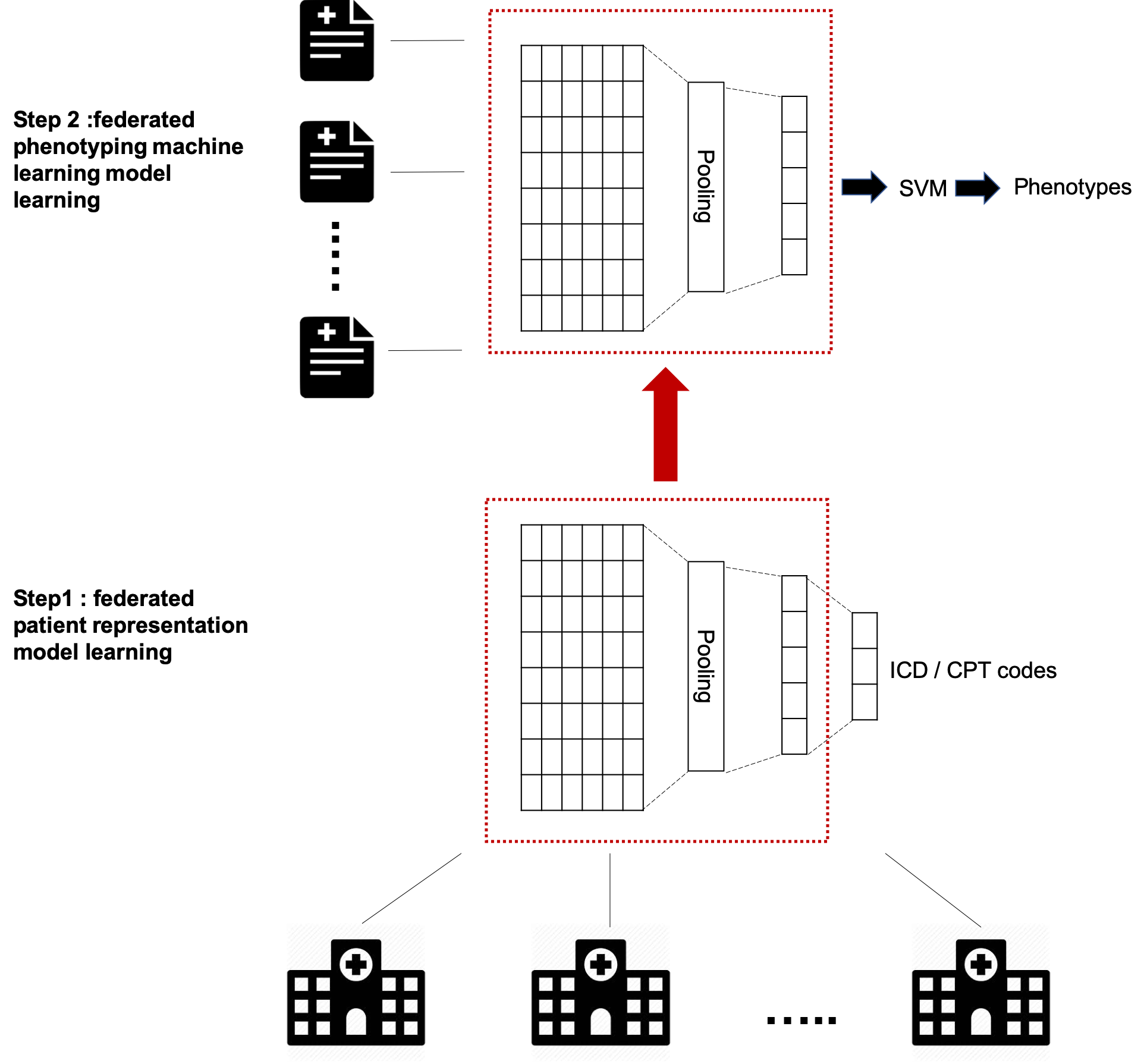} 
  \caption{Two stage federated natural language processing for clinical notes phenotyping. In the first stage, a patient representation model was trained using an artificial neural network (ANN) to predict ICD and CPT codes from the text of the notes from a wide range of healthcare providers. The model without output layer was then used as "representation extractor" in the next stage. In the second stage, a phenotyping support vector machine model was trained in a federated manner using clinical notes for the target phenotype  distributed across multiple silos.}
  \label{fig:Two_stage} 
\end{figure*}

To achieve this goal, we design a two-stage federated NLP approach (Figure \ref{fig:Two_stage}). In the first stage, following \cite{dligach2018learning}, we pre-train a patient representation model by training an artificial neural network (ANN) to predict ICD and CPT codes from the text of the notes. We extend the methods from \cite{dligach2018learning} to facilitate federated training.


In the second stage, a phenotyping machine learning model is trained in a federated manner using clinical notes that are distributed across multiple sites for the target phenotype. In this stage, the notes mapped to fixed-length representations from stage (1) are used as input features and whether the patient has a certain disease is used as a label with one of the three classes: presence, absence or questionable.  

In the following sections, we first describe a simple notes pre-processing step. We then discuss the method for pre-training patient representations and the method for training phenotyping models. Finally, we describe our framework for performing the latter two steps in a federated manner.

\subsection{Pre-processing}

All of our models rely on standardized medical vocabulary automatically extracted from the text of the notes rather than on raw text.

To obtain medically relevant information from clinical notes, Unified Medical Language System (UMLS) concept unique identifiers (CUIs) were  extracted from each note using Apache cTAKES (https://ctakes.apache.org). UMLS is a resource that brings together many health and biomedical vocabularies and standardizes them to enable interoperability between computer systems.

The Metathesaurus is a large, multi-purpose, and multi-lingual vocabulary that contains information about biomedical and health related concepts, their various names, and the relationships among them. The Metathesaurus structure has four layers, Concept Unique Identifies(CUIs), Lexical (term) Unique Identifiers (LUI), String Unique Identifiers (SUI) and Atom Unique Identifiers (AUI). In this study, we focus on CUIs, in which a concept is a medical meaning. Our models use UMLS CUIs as input.

\begin{algorithm*}[t!]
    \caption{Two-stage federated natural language processing }
    \textbf{Stage 1}  
    \BlankLine
      \KwIn{MIMIC3 data clinical notes distributed at 10 simulated sites,
      Representation learning model}
      \BlankLine
      \KwOut{174 ICD or CPT codes}  
      \BlankLine
      Extract CUIs from each patient's clinical notes using cTAKE. 
      \BlankLine
     
      \For{$t\in 1$ \KwTo $T$ }{
      \BlankLine
      \For{$k \in  1$ \KwTo $K$ in parallel  }{
        Train patient representation learning model $f_k$} \
      \BlankLine
      aggregate models from all sites by 
      \(W_{ag}^{t}=\sum_{k=1}^{K} \frac{n_k}{N} w_{k}^{t}\)

        } \;
        
    \textbf{Stage 2}  
    \BlankLine
      \KwIn{i2b2 clinical notes for obesity comorbidities distributed at 3 sites,
      phenotyping machine learning model}
      \BlankLine
      \KwOut{1 single binary output (one of the comorbidities)}  
      \BlankLine
      Extract CUIs from each  clinical notes using cTAKES. 
      \BlankLine
     
      \For{$t\in 1$ \KwTo $T^{'}$ }{
      \BlankLine
      \For{$k \in  1$ \KwTo $K^{'}$ in parallel  }{
        Train phenotyping model $f^{'}_k$} \
      \BlankLine
      aggregate models from all sites by 
      \(W_{ag}^{'t}=\sum_{k=1}^{K^{'}} \frac{n_k^{'}}{N^{'}} w_{k}^{'t}\) 
      
        } \
  
\label{alg.fed}
\end{algorithm*}

\subsection{Representation learning}

We adapted the architecture from \cite{dligach2018learning} for pre-training patient representations. A deep averaging network (DAN) that consists of an embedding layer, an average pooling layer, a dense layer, and multiple sigmoid outputs, where each output corresponds to an ICD or CPT code being predicted. 

This architecture takes CUIs as input and is trained using binary cross-entropy loss function to predict ICD and CPT codes. After the model is trained, the dense layer can be used to represent a patient as follows: the model weights are frozen and the notes of a new patient are fed into the network; the patient representation is collected from the values of the units of the dense layer. Thus, the text of the notes is mapped to a fixed-length vector using a pre-trained deep averaging network.


\subsection{Phenotyping}

A linear kernel Support Vector Machine (SVM) taking input from representations generated using the pre-trained model from stage 1 was used as the classifer for each phenotype of interest. No regularization was used for the SVM and stochastic gradient descent was used as the optimization algorithm.

\subsection{Federated machine learning learning on clinical notes}

 To train the ANN model in either stage 1 or stage 2, we simulated sending out models with identical initial parameters to all sites such as hospitals or clinics. At each site, a model was trained using only data form that site. Only model parameters of the models were then sent back to the analyzer for aggregation but not the original training data. An updated model is generated by averaging the parameters of models distributively trained, weighted by sample size \cite{konevcny2016federated,mcmahan2016communication}. In this study, sample size is defined as the number of patients. 
 
 After model aggregation, the updated model was sent out to all sites again to repeat the global training cycle (Algorithm \ref{alg.fed}). Formally, the weight update is specified by:
 
\begin{equation}
W^{t}_{ag}=\sum_{k=1}^{K} \frac{n_k}{N} W^{t}_{k}
\end{equation}

where $W_{ag}$ is the parameter of aggregated model at the analyzer site, $K$ is the number of data sites, in this study the number of simulated healthcare providers or clinics. $n_i$ is the number of samples at the $i^{th}$ site, $N$ is the total number of samples across all sites, and $W_i$ is the parameters learned from the $i^{th}$ data site alone. $t$ is the global cycle number in the range of [1,T]. The algorithm tries to minimize the following objective function: 

\begin{align*}
\underset{f}{arg \space min} (-\sum_{j=1}^{N}\sum_{p=1}^{M}[y_{jp}log f(x_{jp})+ \\ (1-y_{jp})log(1-f(x_{jp}))])
\label{fig:obj}
\end{align*}

Where $x_{j}$ is the feature vector of CUIs. and $y$ is the class label. $p$ is the output number and $M$ is the total number of outputs. $f$ is the machine learning model such as artificial neural network or SVM.Codes that accompany this article can be found at our github repository \footnote{\url{https://github.com/kaiyuanmifen/FederatedNLP }}. 

\section{Experiments}

To imitate real world medical setting where data are distributed with different healthcare providers, we randomly split patients in MIMIC-III data into 10 sites for stage 1 (federated representation learning).  The training data of i2b2  was split into 3 sites for stage 2 (phenotype learning) to mimic obesity related notes distributed with three different healthcare providers. i2b2 notes were not included in the representation learning as in clinic settings information exchange routes for disease-specific records are often not the same as general medical information and ICD/CPT codes were not available for i2b2 dataset. 
 
Experiments were designed to answer three questions: 

\begin{enumerate}
\item Whether clinical notes distributed in different silos can be utilized for patient representation learning without data sharing
\item Whether utilizing data from a wide range of sources will help improve performance of phenotyping from clinical notes
\item Whether models trained in a two-stage federated manner will have inferior performance to models trained with centralized data.  

\end{enumerate}

To answer these questions, two-stage NLP algorithms were trained. Performance of models trained using only  i2b2 notes from one of the three sites were compared with two-stage federated NLP results. Furthermore, performance of machine learning models using distributed or centralized data at patient representation learning stage or phenotyping stage were compared. 

\section{Results}

\begin{table*}[t!]
\centering 
  \caption{Performance of different experiments} 
\begin{tabular}{|l|l|l|l|l|l|l|}
\hline

\textbf{Experiment} & \textbf{Patient representations} & \textbf{Phenotyping} & \textbf{Precision} & \textbf{Recall} & \textbf{F1} \\ \hline

1 & Bag-of-CUIs & Centralized & 0.649 & 0.627 & 0.634 \\ \hline
2 & Bag-of-CUIs & Federated & 0.650 & 0.623  & 0.632 \\ \hline
3 & Bag-of-CUIs & Single source &  0.552 & 0.540 & 0.542 \\ \hline
4 & Centralized learned & Centralized & 0.749 &  0.714 & 0.726 \\ \hline
5 & Centralized learned & Federated & 0.743 & 0.713 & 0.723 \\ \hline
6 & Federated learned & Centralized & 0.729 & 0.716 & 0.715 \\ \hline
7 & Federated learned & Federated & \textbf{0.753} & \textbf{0.715} & \textbf{0.724} \\ \hline

\end{tabular}
  \label{tab:experiment} 
\end{table*}

\begin{table}[t!]
\centering 
\caption{Performance of two-stage federated NLP in obesity comobidity phenotyping by disease} 
\begin{tabular}{|l|l|l|l|}
\hline
\textbf{Disease}              & \textbf{Prec} & \textbf{Rec} & \textbf{F1} \\ \hline
Asthma               & 0.941              & 0.919           & 0.930       \\ \hline
CAD                  & 0.605              & 0.606           & 0.605       \\ \hline
CHF                  & 0.583              & 0.588           & 0.585       \\ \hline
Depression           & 0.844              & 0.774           & 0.801       \\ \hline
Diabetes             & 0.879              & 0.873           & 0.876       \\ \hline
GERD                 & 0.578              & 0.543           & 0.558       \\ \hline
Gallstones           & 0.775              & 0.619           & 0.650       \\ \hline
Gout                 & 0.948              & 0.929           & 0.938       \\ \hline
Hypercholesterolemia & 0.891              & 0.894           & 0.892       \\ \hline
Hypertension         & 0.877              & 0.854           & 0.865       \\ \hline
Hypertriglyceridemia & 0.725              & 0.519           & 0.524       \\ \hline
OA                   & 0.531              & 0.520           & 0.525       \\ \hline
OSA                 & 0.627              & 0.594           & 0.609       \\ \hline
Obesity              & 0.900              & 0.894           & 0.897       \\ \hline
PVD                  & 0.590              & 0.604           & 0.596       \\ \hline
Venous Insufficiency & 0.763              & 0.712           & 0.734       \\ \hline
\textbf{Average} & 0.753              & 0.715           & 0.724       \\ \hline
\end{tabular}
\label{tab:bycomobidity} 
\end{table}

\subsection{Two-stage federated natural language processing improves performance of automatic phenotyping}

We looked at the scenarios where no representation learning was performed. In those cases, the standard TF-IDF weighted sparse bag-of-CUIs vectors were used to represent i2b2 notes. The sparse vectors were used as input into the phenotyping SVM model. We also looked at the scenarios where representation learning was performed by predicting ICD codes. For each of these conditions, we trained our phenotyping models using centralized vs. federated learning. Finally, we considered a scenario where the phenotyping model was trained using the notes from a single site (the metrics we report were averaged across three sites).

To summarize, seven experiments were conducted: 

\begin{enumerate}

\item No representation learning + centralized phenotyping learning
\item No representation learning + federated phenotyping learning where i2b2 training data were randomly split into 3 silos
\item No representation learning + single source phenotyping learning, where i2b2 data were randomly split into 3 silos, but phenotyping algorithm was only trained using data from one of the silos
\item Centralized representation learning + centralized phenotyping learning 
\item Centralized representation learning + federated phenotyping learning
\item Federated representation learning + centralized phenotyping learning,where MIMIC-III data were randomly split into 10 silos
\item Federated representation learning + federated phenotyping learning, where MIMIC-III data were randomly split into 10 silos and i2b2 data into 3 silos (Table \ref{tab:experiment}).

\end{enumerate}

The results of our experiments are shown in Table \ref{tab:bycomobidity}. First of all, we looked at whether phenotyping model training can be conducted in a federated manner without compromising performance. When only i2b2 data from one of three silos was used for phenotyping training (experiment 3), the F1 score of 0.542 was achieved. When data from all three i2b2 sites were used for phenotyping model training (experiment 1) the F1 score improved to 0.634, which suggests that more data did improve the model.  If we assume data from the three i2b2 silos can not be moved and aggregated together, the model  trained in a federated manner (experiment 2) achieved a comparable F1 score of 0.632. This suggested federated learning worked for phenotyping model training.

Previous work showed that using learned representations from clinical notes from a different source using a transfer learning strategy helps to improve the performance of phenotyping NLP models \cite{dligach2018learning}. When patient representations learned from centralized MIMIC-III notes were used as features and centralized phenotyping training was conducted (experiment 4), the phenotyping performance increased significantly with F1 score of 0.714, which was consistent with previous findings \cite{dligach2018learning}. 

When a federated approach was applied in both representation learning and phenotyping stages, the algorithm achieved F1 score of 0.724. It is worth pointing out that F1 scores from experiment 7 , where both representation and phenotyping training were conducted in a federated manner, were not statistically different from F1 scores of experiment 4 over multiple rounds of experiment using different data shuffling and initialization.  In comparison, when only data from a single simulated silo was used, the average F1 score 0.634. When the centralized approach was taken at both stages, the precision, recall and F1 score were 0.718, 0.711 and 0.714 respectively.  These results suggested  utilizing clinical notes from different silos in a federated manner did improve accuracy of the phenotyping NLP algorithm, and the performance is comparable to NLP trained on centralized data. The performance of federated NLP on each single obesity commodity were shown in Table \ref{tab:bycomobidity}. It is necessary to point out that it was impractical to conduct federated phenotyping training when the number of ``questionable'' cases for many diseases are small (Table \ref{i2b2Cohort}). This is true for many diseases in the i2b2 dataset. In such situation, ``questionable'' cases were excluded from the training and testing process. Instead of 3-class classification, a 2-class binary classification of ``presence'' or ``absence'' were conducted. Therefore, the performance metrics can not be directly compared with results in the original i2b2 challenge, though the scores were similar.





\section{Conclusion} 

In this article, we presented a two-stage method that conducts patient representation learning and obesity comorbidity  phenotyping, both in a federated manner. The experimental results suggest that federated training of machine learning models on distributed datasets does improve performance of NLP on clinical notes compared with algorithms trained on data from a single site. In this study, we used CUIs as input features into machine learning models, but the same federated learning strategies can also be applied to raw text. 
\section{Acknowledgement}
Research reported in this publication was supported by the National Library Of Medicine of the National
Institutes of Health under Award Number R01LM012973. The content is solely the responsibility
of the authors and does not necessarily represent the official views of the National Institutes of
Health.

\bibliography{references,AdditionalCitation}


\bibliographystyle{acl_natbib}

\appendix

\end{document}